# Partisan US News Media Representations of Syrian Refugees

Keyu Chen[1], Marzieh Babaeianjelodar[1], Yiwen Shi[2], Kamila Janmohamed[2], Rupak Sarkar[3], Ingmar Weber[4], Thomas Davidson[5], Munmun De Choudhury[6], Jonathan huang[7], Shweta Yadav[8], Ashique Khudabukhsh[9], Preslav Ivanov Nakov[4], Chris Bauch[10], Orestis Papakyriakopoulos[11], Kaveh Khoshnood[1], and Navin Kumar[1]

[1]Yale University School of Medicine, New Haven, CT 06510, USA
[2]Yale University, New Haven, CT 06510, USA
[3]University of Maryland, College Park
[4]Qatar Computing Research Institute
[5]Rutgers School of Arts and Sciences
[6]Georgia Institute of Technology
[7]Singapore Institute for Clinical Sciences (SICS)
[8]University of Illinois Chicago
[9]Rochester Institute of Technology
[10]the University of Waterloo
[11]Princeton University

## Abstract

We investigate how representations of Syrian refugees (2011-2021) differ across US partisan news outlets. We analyze 47,388 articles from the online US media about Syrian refugees to detail differences in reporting between left- and right-leaning media. We use various NLP techniques to understand these differences. Our polarization and question answering results indicated that left-leaning media tended to represent refugees as child victims, welcome in the US, and right-leaning media cast refugees as Islamic terrorists. We noted similar results with our sentiment and offensive speech scores over time, which detail possibly unfavorable representations of refugees in right-leaning media. A strength of our work is how the different techniques we have applied validate each other. Based on our results, we provide several recommendations. Stakeholders may utilize our findings to intervene around refugee representations, and design communications campaigns that improve the way society sees refugees and possibly aid refugee outcomes.

## Introduction

News media plays a central role in linking individuals to global events (Bleich, Bloemraad, and De Graauw 2015). News media coverage increases the importance of various topics among readers and is representative of some views on the issue (Chandelier et al. 2018). However, news media conveys both facts and varying representations of the same topic, becoming instrumental in individual attitude and public opinion formation (Braha and De Aguiar 2017). Different news media representations are linked with real-world implications. Negative news media representations around marginalized groups may lead to isolation, reduced mental health outcomes, unfairly punitive policy measures that target vulnerable communities (Baranauskas and Drakulich 2018), and hate-crime incidents (Lumsden, Goode, and Black 2019). In this study, within marginalized groups, we focus on Syrian Refugees.

In the last two decades, the number of people displaced worldwide has dramatically increased. The conflict that has engulfed Syria since 2011 has internally displaced 6.5 million people and forced another 5 million to flee abroad, overwhelmingly to neighboring countries, such as Jordan, Lebanon, Turkey, and Iraq (Nassar and Stel 2019). The role of the US regarding the Crisis is central to understanding representations around Syrian refugees. The United States is a major donor to the humanitarian response in Syria, providing humanitarian assistance for vulnerable individuals inside Syria and those displaced in the region since the start of the Syrian Refugee Crisis. However, the US is resettling a relatively small number of Syrian refugees. In 2016, the US had resettled 15,479 Syrian refugees. However, in 2018, 62 refugees were admitted (Romero 2019; Zezima 2019), which is minimal compared to the millions displaced globally. Within the US, public attitudes — and especially partisan attitudes — toward refugees may play an important role in shaping legislators' behavior (Barbera et´ al. 2019), with dire consequences for those seeking refuge from the Syrian Refugee Crisis (hereafter *Crisis*). The UNHCR has thus expressed its concerns on refugee news media representation and its consequences on public opinion (Berry, Garcia-Blanco, and Moore 2018). Public opinion on refugees has become even more polarized, particularly along partisan lines (Adida, Lo, and Platas 2019). For example, in 2017, when Donald Trump was elected in the US, the percentage of Republicans who agreed that the US has a responsibility to accept refugees fell from 35 to 26 percent, while among Democrats, it increased from 71 to 74 percent (Hartig 2018).

Past work used an automated content analysis of Canadian print media coverage over a 10-year period to find that immigrants are represented in economic terms, with an emphasis on the validity of refugee claims, potential security threats, and the extent to which refugees take advantage of social programs (Lawlor and Tolley 2017). Another study examined the representations of Syrian refugees in Canadian print media from 2012 to 2016. Results indicated that the conflict representation was prominent earlier in the Syrian Refugee Crisis but then shifted toward a more humanizing depiction of refugees (Wallace 2018). Other work explored how news outlets discussed refugees, finding that conservative media emphasized refugees as threats more often than liberal media (Nassar 2020). While past work provided an overview of representations around Syrian refugees, there was minimal
focus on the difference in representations between partisan media, using large scale computational techniques, focusing on the larger scope of the Crisis. Thus, we propose a study on US news media representations around Syrian refugees, exploring online news from 2011-2021, focusing on partisan news media outlets (left-leaning, right-leaning, centrist). Better understanding of partisan representations regarding Syrian refugees in the US news media can shed light on how media environments shape partisan views around a vulnerable community, with possible policy implications. We propose the following research question: What are the broad differences between partisan news outlet representations of Syrian refugees?

## Contributions

Past research provided an overview of representations around Syrian refugees using computational techniques. For example, (Brandle and Reilly 2019) conducted content analyses of television news broadcasts, and compared it with US refugee admissions data and data from the United Nations High Commissioner for Refugees. (Adida, Lo, and Platas 2019) conducted a conjoint experiment on a representative sample of 1800 US adults, manipulating refugee attributes in pairs of Syrian refugee profiles, and asked respondents to rate each refugee's appeal. However, these and similar studies did not explore the larger timeline of the Crisis or compare left- and right-leaning representations within this larger period, critical to understanding the evolution of refugee representations, and partisan policies. Our work seeks to fill this gap by exploring 47,388 articles around Syrian refugees, dating from 2011-2021, detailing differences in reporting between left- and right-leaning media. We provide an overview of a decade of articles, expanding on the scope of past work, but more importantly adding understanding around ideological variations in reporting during the period. Previous work indicated that the left-leaning media characterized refugees as victims, compared to right-leaning portrayals of refugees as Islamic terrorists or faceless individuals (Bhatia and Jenks 2018). Similarly, our polarization (KhudaBukhsh et al. 2020) and question answering (Devlin et al. 2018) results tended to represent refugees as child victims, welcome in the US, and right-leaning media cast refugees as Islamic terrorists. We note similar results with our sentiment (Hutto and Gilbert 2014) and offensive speech (Mathew et al. 2021) scores over time, which detail possibly unfavorable representations of refugees in right-leaning media. Based on our results, we provided several recommendations. Stakeholders may utilize our findings to intervene around refugee representations, and design communications campaigns, among other measures, thereby improving representations around Syrian refugees, possibly aiding refugee outcomes, such as decreases in stigma, stress, and poor mental health outcomes experienced by refugees (Henkelmann et al. 2020).

## Background

### Partisan Politics and News Media

A well-functioning democracy requires public discussion and engagement among citizens on key national issues. In the past, this discussion could only be measured by listening to dinner conversations, reading newspaper editorials and political leaflets, or listening to public speeches (Corner 2003). Today, much of the conversation has moved to, and is recorded in, the numerous news articles that appear publicly online daily. We can thus study the effects of the media on the classical notion of expressed public opinion, with a focus not on changes in individual behavior or attitudes but instead on the content of the national conversation. Within analyses of the news media, we note the important role of politicians in agenda setting, as most public opinion is elite-led (Zaller and others 1992). Thus, increasing political polarization can affect agenda setting within the news media (Wilson, Parker, and Feinberg 2020), with possible detrimental consequences. By political polarization, we refer to the degree to which political partisans dislike, distrust, and avoid the other side

(Iyengar et al. 2019). For example, Democrats and Republicans both say that the other party's members are hypocritical, selfish, and closed-minded, and they are unwilling to socialize across party lines, or even to partner with opponents in a variety of other activities (Iyengar et al. 2019). Political polarization in itself is not inimical to a nation's functioning, and there is value in disagreement within a system intended to represent the diverse interests of the electorate. However, profound animosity between parties is a concern. For example, opponent-party animosity may lead parties to promote policy stances more out of disdain for opponents than endorsement for the position. For example, right-leaning politicians in the US may represent Syrian refugees as a threat to the nation (Nassar 2020). Such politicians may propagate their ideas through similarly right-leaning news media, with the aim of promoting their agenda and reaching voters. Consequently, the different partisan representations of news media may play a significant role in the formation of discourse related to Syrian refugees. Our work provides examples of differing partisan representations around refugees in right-leaning media. Findings can facilitate policy interventions that can be harnessed by stakeholders to improve refugee outcomes.

Media, Partisan Representations, and Syrian Refugees

How has the media represented Syrian refugees? First, media coverage of Syrian refugees, and other minorities is not proportional to their actual presence in society (Wright 2002), leading to different representations. Second, news media, especially right-leaning media, through refugee representations, consciously or unconsciously produces and reproduces forms of discrimination, such as stereotypes and prejudice associated with refugees (Brandle and Reilly 2019). Certain representations of refugees in right-leaning news media (Adida, Lo, and Platas 2019) can lead to a distorted picture of the groups. Examples of right-leaning narratives are that refugees are a threat to the American way of life, a burden on national resources, or Islamic terrorists (Bhatia and Jenks 2018). Our findings demonstrate salient examples from the right-leaning media that may increase stigma around refugees. Our work can provide an overview of narratives from which interventions can be designed to mitigate stigma experienced by refugees.

Partisan Media and Real World Effects

These differing partisan representations of refugees with real world implications. Given differing partisan representations within the media, the portrayal of Syrian refugees and the Crisis can perpetuate discrimination toward refugees, especially by individuals consuming only media from a certain partisan affiliation (Consterdine 2018). For the Syrian refugees themselves, negative reporting may further worsen isolation and mental health (Kira et al. 2017). Partisan reporting that has starkly different representations of refugees can lead to the perpetuation of discrimination and othering (Wilmott 2017). Such representations can manifest as directly as racism toward refugees (Ozduzen, Korkut, and Ozduzen 2020). In extreme cases, negative representation of Syrian refugees, or their linkage to terrorist activity, may be related to hate crimes against such communities (Koc¸ 2021). Moreover, while partisan media may represent Syrian refugees in a negative context, this can lead to a general rise of hate crimes targeted against vulnerable communities (Lambert and Githens-Mazer 2010). Similarly, partisan media's portrayal of refugees as an external threat may lead to more restrictive immigration and national security policies (Brandle and Reilly 2019). Results presented in this paper provide evidence on how right-leaning media characterizes refugees as a threat. Given these real world implications, our work can provide understanding on the specific ways partisan news media represents the issues surrounding Syrian refugees, aiding policy interventions.

Data and Method

Data

We used the open-source media analysis platform Media Cloud (mediacloud.org) to analyze 47,388 media articles between 2011 and 2021 from 228 US media sources across the partisan spectrum. For locating articles related to Syrian refugees, we used queries based on a related systematic review (El Arab and Sagbakken 2018): (refugee OR "asylumseeker" OR migrant OR immigrant OR displaced) AND (syria OR syrian). We queried 84,214 URLs (30,401 URLs were broken links) from Media cloud. We did not encounter any issues with scraping text from websites that had a paywall or similar blockers. The distribution of broken links was similar to the distribution of our successfully obtained data, as below, likely indicating that the distribution of missing links is not biased. We categorized these URLs as left, center-left, center, center-right and right, through methods developed in past work (Faris et al. 2017), which used the proportion of retweets associated with either Hillary Clinton or Donald Trump for each media source as a measure of candidate-centric partisanship. The retweets were performed by the Twitter accounts associated with the online news outlets. This metric was expressed on a -1.0 to 1.0 scale. The continuous metric was broken into even quintiles, labelled: left, center-left, center, center-right, and right (Faris et al. 2017). The partisan breakdown of our data was as follows: left: 25.3%, center-left: 37.7%, center: 23.3%, center-right:

2.1%, right: 11.5%. Examples of such sources are as follows: left (Rolling Stone, The Nation); center-left (Fortune, Gawker); center (Forbes, ABC News); center-right (RedState, National Review); right (Breitbart, Blaze). News media sources that appeared after 2016 and were not already categorized were not used in our study. We successfully collected and categorized 47,388 articles. Two reviewers independently examined 100 randomly selected articles to verify salience with our research question. Reviewers then discussed their findings and highlighted items deemed relevant across both lists, to determine that 95% were relevant.

Key Event Selection

As articles around Syrian refugees are contextually embedded within the Syrian Refugee Crisis, we will use key events in the Crisis to inform our descriptive analyses of article count, and sentiment and offensive speech scores. These events will be used to divide the our timeline into five periods. For example, the time between the first event and second event will be Period 1, and the time between the second event and the third event will be Period 2. We developed a list of key events during the Crisis as follows. Three content experts first developed a list of ten events independently based on authoritative sources (UNHCR, UNICEF). Content experts were selected based on their experience around displaced persons, language, and social media analysis. Experts were told to select events based on their relative importance to the Crisis. Experts then compared lists to select common items across lists. There were five events that appeared in all lists, and these were used to demarcate the periods for our descriptive analyses: A-March 2011: Start of unrest in Syria; B-July 2012: Za'atari Refugee Camp open for refugees by UNHCR and Jordanian authorities/first refugee camp in Jordan opens, reaching 100k refugees in its first year; C-January 2016: UNHCR joins humanitarian convoy to deliver life-saving aid to civilians. D-July 2017: In Hamburg, Germany, an agreement is reached on curbing violence in Southwest Syria during G20 meeting, and ceasefire takes effect; E-October 2019: The US withdraws troops from northern Syria and Turkey attacks US Kurdish allies in the area. We note that Event A was selected as it marks the start of the Syrian Refugee Crisis. The following are some events that were evaluated, but not selected for analysis: March 2014: Syrian Army and Hezbollah forces recapture Yabroud, the last rebel stronghold near the Lebanese border; September 2014: US and five Arab countries launch air strikes against Islamic State around Aleppo and Raqqa; July 2015: number of Syrian refugees tops the four million mark; August 2017: UNHCR and partners open Jordan's first job centre for Syrians in Za'atari; October 2019: US withdraws troops from northern Syria.

Offensive Speech

To identify offensive speech in our data, we used BERTHateXplain (Mathew et al. 2021). Offensive speech is strongly impolite, rude or vulgar language expressed towards an individual or group (Davidson et al. 2017). Offensive speech is different from hate speech, which is speech that targets disadvantaged social groups in a manner that is potentially harmful to them (Jacobs and Potter 1998). BERTHateXplain utilizes a large-scale language model, BERT (Devlin et al. 2018), and annotates data from three different perspectives: the commonly used 3-class classification (hate, offensive or normal) (Davidson et al. 2017), the target community (the community that has been the target in the post), and the rationales, or the spans of the post on which the labelling decision is based. BERT-HateXplain was trained on data from Gab and Twitter. In our study, BERT-HateXplain was applied to the first 512 tokens of each article as BERT has a limit of 512 tokens (Devlin et al. 2018). We assumed that key information in the articles would occur in the first 512 tokens. We did not use BERT-HateXplain to identify hate speech as hate speech is minimal on online news sites, across ideological leanings.

Sentiment

We assessed sentiment through VADER (Hutto and Gilbert 2014). VADER is a lexicon and rule-based model for sentiment analysis of text. VADER has been validated by multiple independent human judges (Hutto and Gilbert 2014). VADER presents the following categories for sentiment score: positive sentiment (compound score≥0.05); neutral sentiment (compound score>-0.05 and compound score<0.05); negative sentiment (compound score≤-0.05). The compound score VADER output is the one most commonly used for sentiment analysis (Hutto and Gilbert 2014). The positive, neutral, and negative scores are ratios for proportions of text that fall in each category, and these all add to 1. These metrics are used to analyze the context and presentation of how sentiment is conveyed or embedded in rhetoric for a given sentence. In our results, we compare two sentiment scores or a range of scores and use terms such as *lower sentiment*. As an example, if there are two sentiment scores, score A=0.3 and score B=-0.3, we would say that score B has lower sentiment than score A. This does not necessarily mean that one has less negative or positive sentiment, just that one value is larger than the other. If score A=0.3 and score B=0.1, we still say that score B has lower sentiment than score A. The same logic would apply for a range of scores. For example, if range A=0.3-0.6 and range B=0.1-0.2, we would say that range B has a lower score than range A. To verify sentiment score results, two content experts coded the sentiment of a randomly selected subset of data (1% of total articles) as positive, neutral or negative. Coders demonstrated >85% agreement. A third content expert then reconciled differences between the two coders to produce a final annotated list. The final list was then compared to the VADER output, where we noted 81% agreement, indicating that VADER is sufficient for our analysis. The following are condensed examples of text coded by VADER: neutral (If you're looking for a caffeine buzz first thing in the morning, here's why you should wait an hour or so; The correspondents will be Vice Media co-founder Suroosh Alvi, journalist, documentary filmmaker and author Ben Anderson); positive (A growing number of migrants are finding jobs in Germany, according to data released on Tuesday that will give heart to supporters of Chancellor Angela Merkel's decision to let in hundreds of thousands of war refugees since 2015; Helping refugees pursue degrees and return to their countries should be a priority, say diplomats and observers); negative (in a separate offensive aided by Turkey, have pushed the Islamic State back from the Turkish border and appear to be on the verge of retaking the city; Refugees have been a focal point of political discussion. While campaigning, President-elect Donald Trump has promised to bar Syrian refugees from coming to the United States).

Polarization through Large-scale Language Models

We used large-scale language models (Smith et al. 2017) to understand the differences between left- and right-leaning news articles. Such models can be used for a range of purposes. We used these models to perform single word translation where the model takes a word in a source language as input and outputs an equivalent word in a target language (KhudaBukhsh et al. 2020). For example, in a translation system performing English to Spanish translation, if the input word is hello, the output word will be ola. We apply large-scale language models to partisan news media. All our news articles across partisan media are in English. We build on earlier work (KhudaBukhsh et al. 2020) and treat left- and right-leaning media as two different languages. As our *languages* are actually English from different sub-communities, on most occasions, translations will be identical. As an example, *food* in English used by the left leaning media (leftEnglish) will likely translate into the same in right-English (KhudaBukhsh et al. 2020). The interesting cases are pairs where translations do not match. The output is not inherently misaligned, and the algorithm simply produces word pairs. We determine whether there is a misalignment through human review. Most of the time, pairs will match (aligned). However, sometimes the pairs will not match (misaligned) and this is of interest. An example pair that may not match in our context is *Republican,Democrat*. *Republican* may be used in favorable contexts in right-English, much like how *Democrat* may be used in left-English. Thus, while both words have different meanings and representations in each sub-community, they are treated the same by the translation algorithm, creating a mismatch in translation for *Republican* and *Democrat*. Such word pairs are misaligned pairs. Such mismatches can provide insights on the differences in refugee representations between left- and right-leaning media. We fed the models our all news article data, divided by ideology (left-leaning, right-leaning), as two different languages.

We provide a brief technical overview of the technique used, drawing from (KhudaBukhsh et al. 2020). Let D1 and D2 be two monolingual text corpora authored in languages L1 and L2 respectively. With respect to D1 and D2, V1 and V2 denote the source and target vocabularies. A word translation scheme that translates L1 to L2 takes a source word (W1) as input and produces a single word translation W2 (more details in (KhudaBukhsh et al. 2020)). A translation algorithm (Smith et al. 2017) drives this process. The algorithm requires two monolingual corpora and a bilingual seed lexicon of word translation pairs as inputs. First, two separate monolingual word embeddings are induced using a monolingual word embedding learning model. As per (Smith et al. 2017), FastText (Bojanowski et al. 2017) was used to train monolingual embedding. Next, a bilingual seed lexicon is used to learn an orthogonal transformation matrix, which is then used to align the two vector spaces. Finally, to translate a word from the source language to the target language, we multiply the embedding of the source word with the transformation matrix to align it with the target vector space. Then, the nearest neighbour of the aligned word vector in the target vector space is selected as the translation of the source word in the target language. Two reviewers manually inspected the top 5000 salient translation pairs, ranked by frequency (KhudaBukhsh et al. 2020), between left- and right-leaning media. Reviewers were instructed to independently order the list with most mismatched pairs at the top. By most mismatched we refer to pairs with the greatest difference in meaning, such as *youth, radicals*. Examples of less mismatched pairs are those which are different words but closer in meaning, such as *robust, comprehensive*, and *dramatic, significantly*. The reviewers then compared the top 30 most mismatched pairs in their lists to look for items common to both lists. Eighteen items were common to both lists, and are displayed in the results section. Examples of pairs not selected are *deported, returned*, *radicalized, unfortunately*, and *fraud, unconstitutional*. As a clarification, our goal in using techniques described in (KhudaBukhsh et al. 2020) was not to provide an improvement over an existing technique, but to demonstrate the technique in a different context. While we largely used the work of (KhudaBukhsh et al. 2020) unchanged, we calculated similarity scores between sentences to find illustrative examples of misaligned pairs in left- and right-leaning media where the pairs appear in highly similar contexts - essentially sentences that have similar meanings but with different words. Similarity scores were calculated with Sentence-bert (Reimers and Gurevych 2019), a modification of the pretrained BERT network that uses siamese and triplet network structures to derive semantically meaningful sentence embeddings that can be compared using cosine-similarity.

Question Answering

Question answering can help us to understand how left- and right-leaning media *answer* the same questions about Syrian refugees, perhaps revealing differences in representations around these topics. For example, some partisan media may produce a more inflammatory reply compared to other media, in response to a broad question about Syrian refugees. We used BERT (Devlin et al. 2018) for answer extraction. The model was applied separately on left- and right-leaning articles. Questions were developed based on input from content experts. We selected three content experts who had published at least ten peer-reviewed articles in the last three years around refugee health and safety. Each content expert first developed a list of ten questions separately. The three experts then discussed their lists to result in a final list of four questions that were broadly similar across all three original lists, and final questions are as follows: Why are refugees coming to America? What do you think of refugees? What is happening in Syria? Why are there child refugees? We highlighted one question at a time and fed it to the model.

The model extracts answers for the question leveraging on context information in each article. To stay within the admitted input size of the model, we clipped the length of each article (title + body text) to 512 tokens. Each question provided one answer per article. We randomly sampled 500, 1000, 1500, and 2000 answers per question. We found that a random sample of 1000 answers provided the greatest range and quality of answers, assessed by two reviewers (85% agreement). Range of answers was determined based on the number of different answers provided by each group of answers (500, 1000, 1500, 2000). Quality was determined by the proportion of sensible and non-repetitive answers to each question compared to total number of answers. Sampling 500 answers provided a limited range of answers and few sensible answers to questions. The 1500 and 2000 group of answers had a large range of answers, but many of these tended to be not useful or relevant to the question, such as stopwords. The 1000 answer selection was found by reviewers to have a good range of answers, comparable to the 1500 and 2000 group of answers, but had a greater proportion of sensible answers compared to other groups of answers. We thus randomly sampled 1000 answers per question. Given space limitations and that several answers were repetitive, with numerous non-useful answers, we are unable to present the 1000 answer selection here. Thus, for brevity and clarity, we decided to present a subsample of the 1000 answer selection, providing an overview of answers, without non-useful and repetitive answers. From the sample of 1000 answers, content experts selected the top 5, 10 and 20 most representative answers per question, for both left-, center-, and rightleaning articles. We found that selecting the top 5 most representative answers provided the least repetition and most sensible answers, and thus we present the top 5 answers. Ensuring coders select representative answers rather than stereotypical answers is central to our final results. We first create a list of refugee stereotypes based on past work (Papakyriakopoulos and Zuckerman 2021). We collect refugeerelated stereotypes from the scientific literature, Wikipedia, and Q&A websites. We keep stereotypical words that appear at least in two out of three sources. A third content expert then reviewed the representative answers at every stage, to verify if any answers matched stereotypes in the list. If a match was found, reviewers were told to provide new answers. We planned to instruct reviewers to repeat this process till they had answers not on the stereotype list. However, no reviewer selected answers appearing on our stereotype list, likely due to their content expertise in the area.

## Results

### Overview

We provide an overview of article count, sentiment score, and offensive speech in Figure 1. We provide three-month moving averages for all variables and Akima was used to interpolate missing values (Akima et al. 2016). Article count Figure 1a exhibits similar trends across all media variants.

We note comparable spikes in article count for all media in Period 2 and 3, perhaps indicative of broadly similar

(a) Article count in US online media: 2011-2021

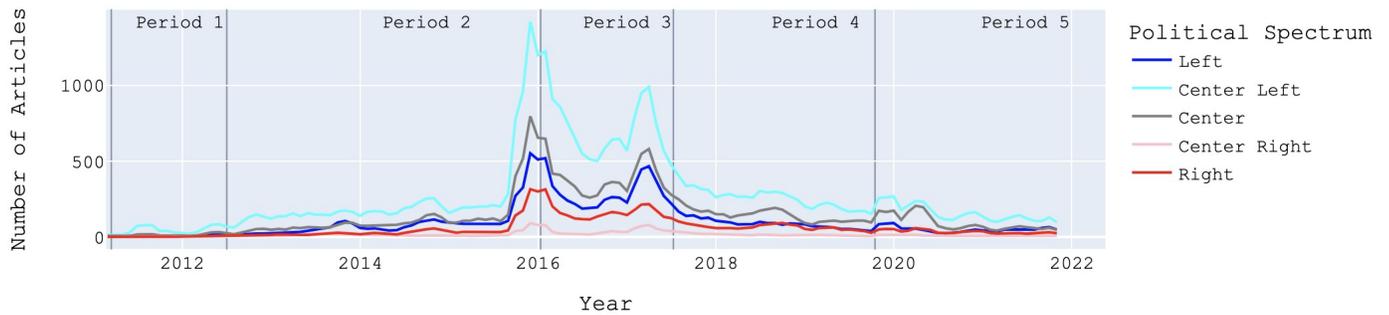

(a)

(b) Sentiment in US online media: 2011-2021

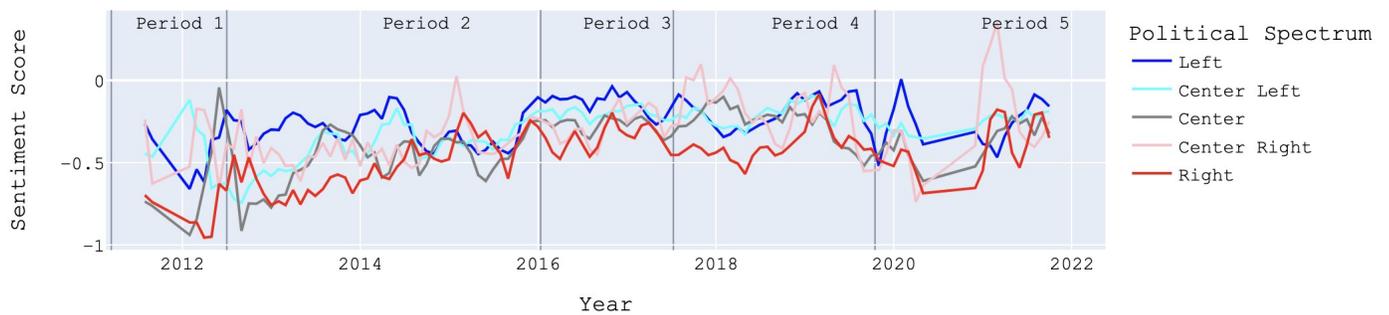

(b)

(c) Offensive speech score in US online media: 2011-2021

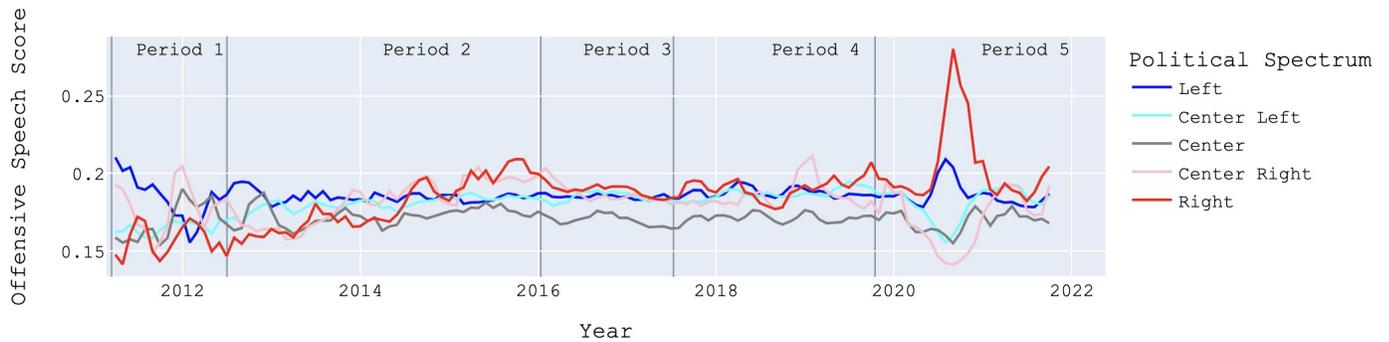

(c)

Figure 1: (a) Article count related to Syrian refugees (2011-2021) across US partisan news outlets online. (b) and (c): Sentiment and Offensive Speech Scores for articles related to Syrian refugees across US partisan online news outlets.

interest in Syrian refugees across media, regardless of ideological leaning. We note that Period 2 is the longest period, and that periods have slightly different lengths. Sentiment score Figure 1b for all media variants generally fluctuated over time. We detail that in Periods 1, 2, 4 and 5, right-leaning media seemed to have lower sentiment compared to other media, perhaps indicating that right-leaning media is less favorable of Syrian refugees compared to other media. Regarding offensive speech Figure 1c, we note that scores fluctuated over our analysis period. Most notably, we note a sharp increase in offensive speech for right-media in Period 5. No similar spikes were observed for any other media throughout the period of study. We suggest that while media across ideological viewpoints has varying representations around Syrian refugees, only right-media promotes a relatively larger amount of offensive speech regarding refugees.

## Polarization

We demonstrate single word translation results from our large-scale language models to understand differences between left- and right-leaning news articles. Upon manual inspection, we present misaligned pairs for left- and rightleaning articles, and illustrative sentence examples in Table 1. We thematically categorized pairs, and then offer illustrative sentence examples within this categorization. We are unable to provide illustrative examples for all pairs due to space constraints.

We first indicate the *conservatives, liberals* pair which demonstrates that the left-leaning media implies that conservatives are highly emotional and irrational, unwilling to accept refugees in the US. Conversely, the right-leaning media implies liberals are easy swayed by refugees looking to exploit the US. Such views may further the partisan divide, affecting overall refugee outcomes. Similarly, the *demonstrators, protesters* pair indicates that the left-leaning media views demonstrators as individuals striving for societal change, unfairly punished by an authoritarian regime. The right-leaning media implies that protesters are responsible for the Syrian conflict, forwarding the view that refugees could be dangerous, ready to incite violence in the US.

We then note the pairs *babies, men*; and *children, people*. It seemed the left-leaning media cast refugees as vulnerable children who needed assistance, rather than men or people who could possibly be terrorists. For context, 47% of Syrian Refugees in the US identify as female, 47% are under the age of 14, and 24% are aged 14-30 (Center 2017). We note the first illustrative sentence for the *babies, men* pair, which indicates that the left-leaning media highlights the struggles refugees have encountered, implying they deserve sympathy and assistance. More importantly, refugees are young and harmless, easy to assimilate into the US social fabric. The right-leaning media details that refugees are opportunistic young men, seeking to move to high-income nations, not deserving of the public's sympathy. Similar results are observed in the second *babies, men* illustrative sentence. The left-leaning media emphasizes that some refugees are babies to be cared for and pitied. However, the right-leaning media implies that many refugees are young men who potentially could be armed and dangerous. Such viewpoints may lead left-leaning members of the public to feel sympathy for helpless refugees, but lead right-leaning individuals to fear and distrust refugees.

We detail the *newcomers, refugees* pair. The left-leaning representation of refugees as newcomers underlies a belief that refugees are welcome in the US, as new immigrants ready to contribute. The right-leaning media uses the generic term refugees instead, perhaps implying that refugees are not immigrants and only in the US temporarily, ready to leave. Such rhetoric may embolden attempts to forcibly expel refugees from the US, affecting refugee well-being. We then note the *demonstrators, protesters* pair. These results may indicate that the left-leaning media characterizes people involved in demonstrations as largely peaceful, compared to the right-leaning media which implies that the same people are violent and dangerous *protesters*. Similar observations were noted for the *DACA, unconstitutional* pair. Deferred Action for Childhood Arrivals (DACA) is a US immigration policy that allows some undocumented individuals in the US after being brought to the country as children to become eligible for US work authorization (Batalova, Hooker, and Capps 2014). The equivalent word in right-leaning media exemplifies right-leaning beliefs that DACA is unconstitutional, further engendering viewpoints that Syrian refugees do not belong in the US.

Compared to the left-leaning media, right-leaning media tended to associate Syrian refugees with Islamic terrorism. For example, while left-leaning media used the word *aliens*, the right-leaning equivalent was *terrorists*. Similar misaligned pairs were *extremism, jihad* and *extremists, islamists*. These examples demonstrate that while left-leaning media does associate Syrian refugees with extremism, the right-leaning media seems to imply that extremism around refugees is linked with Islam. Such viewpoints may lead readers of right-leaning media to believe that Islam is interchangeable with terrorist activity, perhaps increasing stigma towards Muslims. Within these results, a particular concern is the conflation of *islam* with *jihad*, possibly leading consumers of right leaning media to believe Syrian refugees are related to religious extremism. Similarly, we note the *youth, radicals* pair. This pair implies the right-leaning media views youth as a possible danger to the US. By representing refugees as radicals or people, the right-leaning media may lead to people believing that refugees are unable to assimilate into the US and possible radical terrorists. Finally, we note the *cleansing, occupation* pair. Left-leaning media is more likely to represent the Crisis in harsher terms, such as *cleansing*, instead of the slightly less pejorative term *occupation*.

## Question Answering

We present questions and the top five most representative answers per question across left-, right- and center-leaning articles in Table 2. There was a clear difference in characterization of Syrian refugees in left- vs right-leaning articles. Regarding the question *Why are refugees coming to America?* left-leaning media tended to view the Crisis as a great toll on humanity with answers such as *ongoing and brutal situation in war-torn Syria* compared to right-leaning answers which characterized refugees as opportunistic, with answers such as *to get a passport*. Center-leaning media seemed to have mixed answers to the question, with responses such as *a threat to national security* and *refugees enrich communities*.

For the question *What do you think of refugees?*, the leftleaning media provided responses indicating that refugees are innocent individuals who can be easily integrated into the US. Conversely, the right-leaning media views refugees as threats to the US and possible terrorists. Such viewpoints may lead left-leaning members of the public to ascribe sympathy to refugees but incite right-leaning individuals to fear refugees. Exploring center-leaning responses, we observed answers parallel to left- (*supportive*) and right-leaning (*jihadis*) viewpoints

Regarding the question *What is happening in Syria?*, leftleaning media indicates the various traumatic experiences refugees have encountered, allowing the public to sympathize with the refugee cause. The right-leaning media instead views the Crisis as a way for Islamic terrorists to enter high-income European nations, possibly inciting terrorist

| Misaligned Pairs | Left-leaning Illustrative examples | Right-leaning Illustrative examples |
|---|---|---|
| Politics | | |
| <conservatives, liberal> | This week, conservatives are howling over the potential for Syrian refugees to be granted entry into the United States | No matter where the supposed refugees are coming from, liberals want America to take them in |
| <mainstream, liberal> | | |
| <obama, trump> | | |
| <DACA, unconstitutional> | | |
| People | | |
| <babies, men> | So far this year, more than 350 refugees and migrants have drowned on the crossing from Turkey, including women, children, and babies | But the overwhelming number of "refugees" are young men leaving the safety of Turkey, Jordan and other states in the hopes they'll enjoy the wonders of Europe |
| | Images of Syrian refugees clutching their babies on the trek to asylum struck a chord in the hearts of some Minnesota women who met Wednesday to help a nonprofit group lighten the migrants' load | Who are the refugees? While many are women and children, there are plenty of young men of military age? Look in the background of the many photos of the refugees, such as these in the New York Times |
| <children,people> | | |
| <demonstrators,protesters> | | |
| <newcomers, refugees> | | |
| Extremism | | |
| <extremism, jihad> | Islamic State fighters are increasing in Libya, raising concerns that the country could be the next battleground | The number of jihadists in Libya linked to the Islamic State, also known as ISIS or ISIL, |
| | for extremism, and terrorist activities | has been growing in recent months |
| <extremists, islamists> | When you travel to the region, and you hear extremists in the region saying, "America hates us, hates Muslims," and you try to explain [that this isn't the case], this is now overshadowed by the Trump rhetoric | At the same time, it is ridiculous to not recognize there are radical islamists who are in America, who want to bring this country down and who think they go to paradise if they kill Americans |
| <youth, radicals> | She noted the threat to France of the Islamic State group which has claimed deadly attacks in Paris, Nice and elsewhere, and has lured hundreds of French youths to the war zones in Syria and Iraq | French authorities are particularly concerned about the threat from hundreds of French Islamic radicals who have traveled to Syria and returned home, possibly with dangerous skills |
| <islam, jihad> | | |
| <aliens, terrorists> | | |
| <nationalist, islam> | | |
| War | | |
| <cleansing, occupation> | "Turkey's military operation in northern Syria, spearheaded by armed Islamist groups on its payroll, represents an intentioned-laced effort at ethnic cleansing," | A senior Syrian Kurdish official says Turkey's offensive on the Syrian town of Afrin is an "occupation" that endangers the rest of northern Syria |
| | "There was no chance Erdogan would keep his promise, and full blown ethnic cleansing is underway by Turkish supported militias," he said | "The statement from Erdogan's office insisted Turkey "has no interest in occupation or changing demographics" and accused the PKK and YPG of already making efforts to do so |
| <peace, compromise> | | |
| <war, invasion> | | |
| <migration, crisis> | | |

Table 1: Misaligned word pairs and illustrative sentence examples for left- and right-leaning media regarding refugee representations from 2011 - 2021. We are unable to provide illustrative examples for all pairs due to space constraints. activity. The opposing views on reasons for the Crisis may lead left-leaning individuals to empathize, but persuade the right-leaning public into believing the Crisis is a way for Islamic terrorism to infiltrate the US. We detail center-leaning answers to provide context, where responses mirrored both left- (*persecution and conflict*) and right-leaning (*the rise of ISIS*) viewpoints.

Similarly, for the question *Why are there child refugees?*, left-leaning media represents refugees as *victims of war...* or *in search of a better future*. Right-leaning media tended to use discriminatory rhetoric, with responses such as *killing innocent Americans* and *plotting to join Jihadi...*. Centerleaning articles demonstrated a mix of positive and negative inclinations, unlike left- and right-leaning media which had a clear stance toward Syrian refugees. For example, for the question *Why are refugees coming to America?*, centerleaning media had both positive (*refugees enrich communities*) and negative (*a threat to national security*) views on the Crisis and Syrian refugees.

| Why are refugees coming to America? | What do you think of refugees? | What is happening in Syria? | Why are there child refugees? |
|---|---|---|---|
| Left-leaning | | | |
| ongoing and brutal situation in war-torn Syria | new citizens | mass killings | victims of war and religious persecutions |
| poverty, neglect, and violence | innocent | suffocation of refugees | push of war, famine, and poverty |
| The majority of Syrian refugees are fleeing their brutal government | civilians | internal conflict, refugees and displaced people | fleeing civil war in Syria |
| to help relatives pinned down by the violence to escape to safer ground | They feel so abandoned by the world | enslavement and rape | to escape the security forces' violence |
| we begin our stories with victimhood and end them with survival | you are not alone | Stop ISIS terrorists now, before it's too late | in search of a better future |
| Right-leaning | | | |
| to get a passport | unclean | Terrorists have struck in the streets and subways | killing innocent Americans |
| real national security threats are costing American lives | mass violence | real national security threats are costing American lives | plotting to join jihadi fighters in Syria |
| foreign relatives living in countries designated as state-sponsors of terrorism | dominant asymmetric threat to our national security | mass refugee crisis | migrant attacks |
| the country is paying a steep price for putting out the welcome mat | a disgrace to humanity | illegal migrants from smuggler boats and ferrying them to Europe | People are coming in and we know what we're going to have problems |
| they should be deported post-haste | Homegrown terrorists | the migrant crisis to smuggle jihadis in to the United Kingdom | to give Trump some wall money in exchange for protecting the Dreamers |
| Center-leaning | | | |
| brutality of ISIS | poorest | Syrian crisis | struggling with life |
| seeking better living conditions | illegal | the rise of ISIS | trying to enter the country |
| a threat to national security | jihadis | mass refugee crisis | migrant attacks |
| total destruction | there are no fundamentalists | persecution and conflict | they have relatives |
| refugees enrich communities | supportive | jihadist attacks | they want to kill us |

Table 2: Illustrative examples of questions and answers for left-, center-, and right-leaning articles.

## Discussion

### Implications of Findings

Our RQ was to explore the broad differences between partisan news outlets in regard to Syrian refugees. A strength of our work is how the different techniques we have applied validate each other. For example, the sentiment and offensive speech scores over time detail possibly unfavorable representations of refugees in right-leaning media. Similarly, our polarization and question answering results both indicated that the left-leaning media tended to represent refugees as child victims, welcome in the US, and right-leaning media cast refugees as Islamic terrorists. The concordance in our results suggests the veracity of our findings and we hope that results can add to research and policy around refugees and other displaced individuals. Our evidence is supported by previous research. Past work indicated that left-leaning US media often casts refugees as victims or individuals who can someday contribute to the US, allowing people to sympathize with refugees (Bhatia and Jenks 2018). Conversely, right-leaning media represents refugees as a burden or threat to the US (Bhatia and Jenks 2018). However, previous work does not explore the sheer range of media articles around Syrian refugees using NLP techniques. We expand on previous research, providing a media overview of Syrian refugee representations and contrasts between left- and right-leaning media.

### Recommendations

Key to refugee representations is the inclusion of refugee viewpoints when reporting on Syrian refugees. Where possible, refugees themselves should be consulted on news articles about Syrian refugees, co-creating work (Mitchell 2019). For example, journalists can submit articles to a panel staffed by refugees who will then provide suggestions on how the article can represent refugee concerns. To improve representations around refugees, minimize marginalization, and possibly mitigate effects of the Crisis, government stakeholders can conduct tailored interventions and communications campaigns to counter the possibly negative media rhetoric. An example intervention may use stories around extended contact with refugees to build a common ingroup identity among refugees and other individuals in right-leaning areas of the US (Cameron et al. 2006). Such interventions may shift the beliefs of right-leaning Americans around refugees, thereby improving feelings of inclusiveness among refugees with possible implications for refugee mental health (Correa-Velez, Gifford, and Barnett 2010). Communications campaigns can harness our findings and forward evidenced-based posts about refugees in the comment sections of right-leaning online media. Such campaigns may allow those antagonistic to refugees to engage with evidence-based information, perhaps shifting their views on refugees. We also suggest that the media, especially right-leaning media, be more aware of possibly offensive language, to avoid further marginalizing refugees. News organizations can strengthen internal review procedures to ensure they do not use offensive language. Such procedures may improve opinion around refugees, perhaps promoting acceptance towards refugees. More balanced reporting may contribute to the integration of refugees and other vulnerable social groups into US society, perhaps also reducing hatecrime incidents (Papakyriakopoulos and Zuckerman 2021). These recommendations may enhance efforts to integrate refugees and provide a more inclusive environment for them, perhaps mitigating effects of the Crisis.

Given we suggest recommendations around government media interventions and more responsible reporting, we provide further discussion in these areas. Some positions suggest that any government regulation is contrary to protections around free speech and violates US First Amendment rights (Samples 2019). Other viewpoints suggest that governments have a duty to prohibit hateful speech, while not extending such oversight into abusing their authority to silence peaceful dissent (Amnesty International 2020). Our findings lend support to broad media freedoms, but recommend government intervention around hateful content. In this regard, we suggest safeguards to ensure that government intervention around hateful media does not restrict freedom of speech. We note that there also exists the question if the media should be a responsible producer of news e.g., should media outlets be responsible providers of news and not promote hateful or discriminatory content? Some suggest that the media should report as they wish, and it is up to the reader to decide the veracity of content (Meehan 2020). Other viewpoints indicate that the media should be held accountable and minimize harm to those it reports on (Walsh 2016). Our results suggest that the media should avoid stigmatizing vulnerable communities and promoting messaging harmful to marginalized groups, leading to recommendations above, where right-leaning media can limit its discriminatory tone toward refugees.

Limitations

Our findings relied on the validity of data collected with our search terms. We used Media Cloud to search for all articles relevant to Syrian refugees, and our data contained text representative of refugee representations. We are thus confident in the comprehensiveness of our data. We note that the recall of the search string was not tested, and that there may be possible biases as we did not manage to scrape all URLs due to broken links. Our data may not be generalizable to US representations around non-Syrian refugees (Han and Anderson 2021). In future, we will expand our study to broader refugee communities. We were not able to obtain read or share counts, to control for news outlets that are more widely read compared to a small town newspaper. We were not able to distinguish between *bias free* publications and opinion/commentary articles. The team was unable to conduct more fine-grained analysis, e.g. are there news outlets whose representation of refugees has changed? Future research will incorporate such analysis. Findings may also not apply to other related events that are also heavily politicized (e.g., migration from Mexico and Central American) or other contexts (e.g., the experience of the Syrian Refugee Crisis in Europe). Future work will take a broader approach, incorporating other crises. We also note limitations of BERT, such as its inability to learn in few-shot settings (Tanzer,¨ Ruder, and Rei 2021). Such model limitations hampered our ability in analyzing representations around subgroups of refugees, such as LGBT+ refugees, around which there were relatively few news articles. Similarly, due to relatively low numbers of far-right articles, we were unable to conduct analysis of this subset with BERT.

## Conclusion

Broadly, our findings both indicated that the left-leaning media tended to represent refugees as child victims, and equivalent text in right-leaning media cast refugees simply as people or Islamic terrorists. Stakeholders may utilize our findings to intervene around refugee representations, and design communications campaigns, among other measures, thereby improving representations around Syrian refugees, possibly aiding refugee outcomes, such as decreases in stigma, stress, and poor mental health outcomes experienced by refugees.

## Broader perspective, ethics and competing interests

Possible positive outcomes of our work include: more balanced media reporting, reduction in stigma around refugees. Negative outcomes include: media interpreting this study as government propaganda and increasing their negative tone toward refugees. We consulted with stakeholders working with refugees to ensure the work does not further stigmatize vulnerable populations. We report no competing interests.

## References

Adida, C. L.; Lo, A.; and Platas, M. R. 2019. Americans Preferred Syrian Refugees Who are Female, English-speaking, and Christian on the Eve Of Donald Trump's Election. *PloS one* 14(10):e0222504.

Akima, H.; Gebhardt, A.; Petzold, T.; Maechler, M.; et al. 2016. Package 'akima'. *Version 0.6* 2.

Amnesty International. 2020. Freedom of Expression.

Baranauskas, A. J., and Drakulich, K. M. 2018. Media Construction of Crime Revisited: Media Types, Consumer Contexts, and Frames of Crime and Justice. *Criminology* 56(4):679–714.

Barbera, P.; Casas, A.; Nagler, J.; Egan, P. J.; Bonneau, R.;´ Jost, J. T.; and Tucker, J. A. 2019. Who Leads? Who Follows? Measuring Issue Attention and Agenda Setting by Legislators and the Mass Public Using Social Media Data. *American Political Science Review* 113(4):883–901.

Batalova, J.; Hooker, S.; and Capps, R. 2014. DACA at the Two-year Mark: A National and State Profile of Youth Eligible and Applying for Deferred Action. *Migration Policy Institute*.

Berry, M.; Garcia-Blanco, I.; and Moore, K. 2018. Press Coverage of the Refugee and Migrant Crisis in the EU: A Content Analysis of Five European Countries. United Nations High Commissioner for Refugees; 2016.

Bhatia, A., and Jenks, C. J. 2018. Fabricating the American Dream in US Media Portrayals of Syrian Refugees: A Discourse Analytical Study. *Discourse & Communication* 12(3):221–239.

Bleich, E.; Bloemraad, I.; and De Graauw, E. 2015. Migrants, Minorities and the Media: Information, Representations and Participation in the Public Sphere. *Journal of Ethnic And Migration Studies* 41(6):857–873.

Bojanowski, P.; Grave, E.; Joulin, A.; and Mikolov, T. 2017. Enriching Word Vectors with Subword Information. *Transactions of the Association for Computational Linguistics* 5:135–146.

Braha, D., and De Aguiar, M. A. 2017. Voting Contagion: Modeling and Analysis of a Century of US Presidential Elections. *PloS one* 12(5):e0177970.

Brandle, S. M., and Reilly, J. E. 2019. Seldom, Superficial, and Soon Gone: Television News Coverage of Refugees in the United States, 2006–2015. *Refugee Survey Quarterly* 38(2):159–194.

Cameron, L.; Rutland, A.; Brown, R.; and Douch, R. 2006. Changing Children's Intergroup Attitudes Toward Refugees: Testing Different Models of Extended Contact. *Child Development* 77(5):1208–1219.

Center, R. P. 2017. Department of State Bureau of Population, Refugees, and Migration Office of AdmissionsRefugee Processing Center Summary of Refugee Admissions.

Chandelier, M.; Steuckardt, A.; Mathevet, R.; Diwersy, S.; and Gimenez, O. 2018. Content Analysis of Newspaper Coverage of Wolf Recolonization in France Using Structural Topic Modeling. *Biological Conservation* 220:254–261.

Consterdine, E. 2018. State-of-the-art Report on Public Attitudes, Political Discourses and Media Coverage on the Arrival of Refugees. *Ceaseval Research On The Common European Asylum System (02)*.

Corner, J. 2003. *Soft News Goes to War: Public Opinion and American Foreign Policy in the New Media Age*. Princeton University Press.

Correa-Velez, I.; Gifford, S. M.; and Barnett, A. G. 2010. Longing to Belong: Social Inclusion and Wellbeing Among Youth With Refugee Backgrounds In The First Three Years In Melbourne, Australia. *Social Science & Medicine* 71(8):1399–1408.

Davidson, T.; Warmsley, D.; Macy, M.; and Weber, I. 2017. Automated Hate Speech Detection and the Problem of Offensive Language. In *Proceedings of the International AAAI Conference on Web and Social Media*, volume 11.

Devlin, J.; Chang, M.-W.; Lee, K.; and Toutanova, K. 2018. Bert: Pre-training of Deep Bidirectional Transformers for Language Understanding. *arXiv preprint arXiv:1810.04805*.

El Arab, R., and Sagbakken, M. 2018. Healthcare Services for Syrian Refugees in Jordan: A Systematic Review. *European journal of public health* 28(6):1079–1087.

Faris, R.; Roberts, H.; Etling, B.; Bourassa, N.; Zuckerman, E.; and Benkler, Y. 2017. Partisanship, Propaganda, and Disinformation: Online Media and the 2016 US Presidential Election. *Berkman Klein Center Research Publication* 6.

Han, S., and Anderson, C. K. 2021. Web Scraping for Hospitality Research: Overview, Opportunities, and Implications. *Cornell Hospitality Quarterly* 62(1):89–104.

Hartig, H. 2018. Republicans Turn More Negative Toward Refugees as Number Admitted to US Plummets. *Pew Research Center*.

Henkelmann, J.-R.; de Best, S.; Deckers, C.; Jensen, K.; Shahab, M.; Elzinga, B.; and Molendijk, M. 2020. Anxiety, Depression and Post-traumatic Stress Disorder in Refugees Resettling in High-income Countries: Systematic Review and Meta-analysis. *BJPsych open* 6(4).

Hutto, C., and Gilbert, E. 2014. Vader: A Parsimonious Rule-based Model for Sentiment Analysis of Social Media Text. In *Proceedings of the International AAAI Conference on Web and Social Media*, volume 8.

Iyengar, S.; Lelkes, Y.; Levendusky, M.; Malhotra, N.; and Westwood, S. J. 2019. The Origins and Consequences of Affective Polarization in the United States. *Annual Review of Political Science* 22:129–146.

Jacobs, J. B., and Potter, K. 1998. *Hate Crimes: Criminal law & identity Politics*. Oxford University Press on Demand.

KhudaBukhsh, A. R.; Sarkar, R.; Kamlet, M. S.; and Mitchell, T. M. 2020. We Don't Speak the Same Language: Interpreting Polarization through Machine Translation. *arXiv preprint arXiv:2010.02339*.

Kira, I. A.; Shuwiekh, H.; Rice, K.; Al Ibraheem, B.; and Aljakoub, J. 2017. A Threatened Identity: The Mental Health Status of Syrian Refugees in Egypt and its Etiology. *Identity* 17(3):176–190.

Koc¸, G. 2021. Derrida's Concept of Hos(ti)pitality and Hate Crime in Turkey. *Journal of Business Diversity* 21(2).

Lambert, R., and Githens-Mazer, J. 2010. Islamophobia and Anti-Muslim Hate Crime: UK Case Studies. *Islamophobia and Anti-Muslim* 115.

Lawlor, A., and Tolley, E. 2017. Deciding Who's Legitimate: News Media Framing of Immigrants and Refugees. *International Journal of Communication* 11:25.

Lumsden, K.; Goode, J.; and Black, A. 2019. 'I Will not be Thrown out of the Country Because I'm an Immigrant': Eastern European Migrants' Responses to Hate Crime in a Semi-rural Context in the Wake of Brexit. *Sociological Research Online* 24(2):167–184.

Mathew, B.; Saha, P.; Yimam, S. M.; Biemann, C.; Goyal, P.; and Mukherjee, A. 2021. HateXplain: A Benchmark Dataset for Explainable Hate Speech Detection. *Proceedings of the AAAI Conference on Artificial Intelligence* 35(17):14867–14875.

Meehan, D. 2020. Responsibilities of the Media: Are Journalist's Responsible for Publishing the Truth?

Mitchell, M. 2019. *Making Media with Refugee Youth in the UK and Lebanon: A Practice-based Enquiry Into Cocreation*. Ph.D. Dissertation, Royal Holloway, University of London.

Nassar, J., and Stel, N. 2019. Lebanon's Response to the Syrian Refugee Crisis–Institutional Ambiguity as a Governance Strategy. *Political Geography* 70:44–54.

Nassar, R. 2020. Framing Refugees: The Impact of Religious Frames on US Partisans and Consumers of Cable News Media. *Political communication* 37(5):593–611.

Ozduzen, O.; Korkut, U.; and Ozduzen, C. 2020. 'Refugees are not Welcome': Digital Racism, Online Place-making and the Evolving Categorization of Syrians in Turkey. *New Media & Society* 1461444820956341.

Papakyriakopoulos, O., and Zuckerman, E. 2021. The Media During the Rise of Trump: Identity Politics, Immigration,"Mexican" Demonization and Hate-Crime. In *Proceedings of the International AAAI Conference on Web and Social Media*, volume 15, 467–478.

Reimers, N., and Gurevych, I. 2019. Sentence-bert: Sentence Embeddings Using Siamese Bert-networks. *arXiv preprint arXiv:1908.10084*.

Romero, C. 2019. Analyzing the United States' Limited Response to the Syrian Refugee Crisis. *Political Analysis* 20(1):3.

Samples, J. 2019. Why the Government Should Not Regulate Content Moderation of Social Media. *Cato Institute Policy Analysis* 865.

Smith, S. L.; Turban, D. H.; Hamblin, S.; and Hammerla, N. Y. 2017. Offline Bilingual Word Vectors, Orthogonal Transformations and the Inverted Softmax. *arXiv preprint arXiv:1702.03859*.

Tanzer, M.; Ruder, S.; and Rei, M. 2021. BERT Memorisa-¨ tion and Pitfalls in Low-resource Scenarios. *arXiv preprint arXiv:2105.00828*.

Wallace, R. 2018. Contextualizing the Crisis: The Framing of Syrian Refugees in Canadian Print Media. *Canadian Journal of Political Science/Revue Canadienne De Science Politique* 51(2):207–231.

Walsh, L. 2016. Commentary: Media Have Responsibility to the Public. Section: Opinion, opinion, opinion.

Wilmott, A. C. 2017. The Politics of Photography: Visual Depictions of Syrian Refugees in UK Online Media. *Visual Communication Quarterly* 24(2):67–82.

Wilson, A. E.; Parker, V.; and Feinberg, M. 2020. Polarization in the Contemporary Political and Media Landscape. *Current Opinion in Behavioral Sciences* 34:223–228.

Wright, T. 2002. Collateral Coverage: Media Images of Afghan Refugees During the 2001 Emergency.

Zaller, J. R., et al. 1992. *The Nature and Origins of Mass Opinion*. Cambridge University Press.

Zezima, K. 2019. The US has Slashed its Refugee Intake. Syrians Fleeing War are Most Affected. *The Washington Post*.